\documentclass[pre,showpacs,twocolumn]{revtex4}
\usepackage{amsmath,amssymb,graphicx}
\usepackage{color,ulem}

\begin{document}
\newcommand*{\wfnone}{%
\mathrel{\hbox{$\bullet$}\hskip-0.7ex\vcenter{\hbox{$\diagup$}\vskip-1.2ex\hbox{$\diagdown$}}\hskip-0.7ex 
\vcenter{\hbox{$\bullet$}\vskip-1.2ex\hbox{\hskip.2ex$|$}\vskip-1.2ex\hbox{$\bullet$}}}}

\newcommand*{\wfntwo}{\mathrel{\hspace{0.2em}\hbox{$\bullet$}\hspace{-0.2em}\hbox{$-$}\hspace{-0.2em}\hbox{$\bullet$}}}

\newcommand*{\wfnthree}{%
\mathrel{\hbox{$\bullet$}\hskip-0.7ex\vcenter{\hbox{$\diagup$}\vskip-1.2ex\hbox{$\diagdown$}}\hskip-0.8ex 
\vcenter{\hbox{$\bullet$}\hbox{$\bullet$}}}}

\newcommand*{\wfnfour}{%
\mathrel{\hbox{$\bullet$}\hskip-0.7ex\vcenter{\hbox{$\diagup$}\vskip-1.2ex\hbox{$\diagdown$}}\hskip-0.7ex 
\vcenter{\hbox{$\bullet$}\vskip-1.2ex\hbox{\hskip.2ex$|$}\vskip-1.2ex\hbox{$\bullet$}}\hskip-0.7ex
\vcenter{\hbox{$\diagdown$}\vskip-1.2ex\hbox{$\diagup$}}\hskip-0.9ex\hbox{$\bullet$}
}}

\title{Stochastic Weighted Fractal Networks}
\author{Timoteo Carletti} 
\affiliation{D\'epartement de Math\'ematique, Facult\'es Universitaires
  Notre Dame de la Paix\\ 8 rempart de la vierge B5000 Namur, Belgium\\
timoteo.carletti@fundp.ac.be}
\date{\today}

\begin{abstract}
In this paper we introduce new models of complex weighted
    networks sharing several properties with fractal sets: the deterministic 
non-homogeneous 
    weighted fractal networks and the stochastic weighted fractal
    networks. Networks of both classes can be 
  completely analytically characterized in terms of the involved
  parameters. The proposed algorithms improve and extend the framework of
  weighted fractal networks recently proposed in~\cite{carlettirighi}.
\end{abstract}

\maketitle

\section{Introduction}
\label{sec:intro}

Fractal structures are ubiquitous in nature, coastlines~\cite{mandelbrot1967},
river networks~\cite{river,river2}, snowflakes~\cite{NittmanStanley1987},
growing colonies of
bacteria~\cite{Matsuyama1989,FujikawaMatsushita1989,FujikawaMatsushita1991},
mammalian lungs~\cite{SBHPS1994,BBSS1996,KS1997,AAAMBZSS1998,KT2007},
mammalian bloody vessels~\cite{KT2007}, just to mention few of
them~\footnote{The 
  interested reader could find many more example in the beautiful
  books~\cite{vicsek1992,librofisica}.}. But  
also mankind artifacts can 
exhibit fractal features, for instance fractal antenna~\cite{HohlfeldCohen1999}
or fluctuations in markets prices~\cite{mandelbrot1963}.  

A distinction can be made between {{\it mathematical} or deterministic
fractals~\cite{Mandelbrot1982} for which a complete geometric description can
be provided using simple tools such as homotheties, rotations and copying, and
{\it random} or {\it pseudo} fractals~\cite{vicsek1992} found in nature, 
being the
latter characterized by exhibiting fractal properties, for instance
self--similarity, only when
statistical averages are computed, because unavoidable fluctuations and errors
can alter the regular--geometric patterns. Moreover such scale invariance
should be limited to a finite range of scale lengths because of physical
constraints. 

It is worth remarking that 
some of these physical fractals have functionalities, 
e.g. transportation of gases in mammalian lungs, or charges in fractal
antenna, one can thus improve the
geometrical description by including flows and growths constraints. Networks are
therefore the most natural and useful tool to describe such growing complex
structures with flows constraints. We thus hereby propose the {\it Stochastic Weighted Fractal Networks}, SWFN for short, a new class of complex networks whose construction is directly
inspired by such physical fractal structures.

Starting from the pioneering works of Erd\H os and R\'enyi~\cite{ErdosReny1959}, network theory is nowadays a research  
field in its own~\cite{AB2002,BLMCH2006} and the scientific activity is mainly
devoted to construct and characterize complex networks exhibiting some of the
remarkable properties of real networks, scale--free~\cite{BA1999},
small--world~\cite{WattsStrogatz1998}, communities~\cite{Fortunato2009},
weighted links~\cite{YJB2001,ZTZH2003,DM2004,BBPV2004,BBV2004,BBV2004b}, just to mention few of them.  

In recent years we observed an increasing number of papers were authors proposed
models of deterministic (pseudo) fractal networks~\cite{BRV2001,JKK2002,DGM2002,RB2003,ZZCGFZ2007,Zhang2008,Guan2009,ZZZG2009,ZZZCG2007,carlettirighi} exhibiting scale-free and hierarchical structures. In a limited number of cases, models presented also a stochastic component~\cite{ZZSZG2008,WDS2007,CHLR2003,WDDS2006}.

The aim of the SWFN hereby introduced, is to provide a framework that could be used to (re)analyze flows on natural fractal structures using standard tools of transport theory on networks. Moreover SWFN share with physical fractals several interesting properties, for instance the self-similarity or the self-affinity, the presence of hierarchical structures and a stochastic growth process. Actually this allows us to generalize in a unifying scheme some of the above mentioned models existing in the literature.

The SWFN are constructed via a stochastic process and we are thus 
able to analytically characterize their topology as a function of
the parameters involved in the construction, using {\it expectations} obtained
constructing several replicas.

Let us conclude this introduction with two remarks. First of all we named our 
models \lq\lq fractal\rq\rq networks instead of \lq\lq pseudo fractal\rq\rq, 
because some of the topological properties of SWFN depend on the fractal 
dimension of some underlying fractal set, whose value ranges all the positive
 real numbers, without any limitation. Second we rather 
prefer to talk about \lq\lq stochastic\rq\rq networks to emphasize the 
stochastic growth process instead of the randomness of some topological quantities; let us also stress that in the network theory \rq\rq randomness\rq\rq has a precise meaning that we cannot directly apply to this case.

The paper is organized as follows. In the next section we will introduce and
study a deterministic model, that generalize the one proposed
in~\cite{carlettirighi}, and that will serve as the basic building block to
construct the SWFN in 
Section~\ref{sect:StochWFN}. Then we conclude with some possible applications
we sum up and draw our conclusions 


\section{Deterministic Weighted Fractal Network}
\label{sect:nHWFN}

 According to Mandelbrot~\cite{Mandelbrot1982} \lq\lq a fractal is by
 definition a set for which the Hausdorff dimension strictly
 exceeds the topological dimension\rq\rq. One of the most amazing and
 interesting feature of fractals is their {\it self-similarity} or {\it
   self-affinity}~\cite{Mandelbrot1985,Mandelbrot1986}, namely looking
 at all scales we can find conformal or stretched copies of the whole set;
 this is actually the idea used to build up fractals as fixed point of {\it
   Iterated Function Systems}~\cite{barnsley1988,edgar1990}, IFS for
 short. Such fractals have a Hausdorff dimension completely characterized by
 the number of copies and the scaling factors of the IFS. Let us observe that 
 in this case this dimension coincides with the so  called similarity
 dimension~\cite{edgar1990}. 

Recently, author proposed~\cite{carlettirighi} a new general framework aiming to
construct weighted networks with some a priori prescribed topology depending
on the two main parameters: the number of copies and the scaling
factors, hence on the fractal dimension of the \lq\lq underlying\rq\rq IFS
fractal. The aim of this Section is to generalize such construction to obtain 
a larger class of networks; moreover exploiting
 the iterative construction we will be able to completely and analytically
 describe the network topology in terms of node strength distribution, average
 (weighted) shortest path and (weighted) clustering coefficient. 

Let us fix a positive integer $s>1$ and $s$ real numbers $f_1,\dots ,f_s\in
(0,1)$ and let us consider a (possibly) weighted network $G$ composed by $N$ 
nodes, one of which has been labeled {\it attaching node} and denoted by
$a$. We then introduce a map,
$\mathcal{T}_{s,\mathbf{f},a}$, depending on the parameters $s$,
$\mathbf{f}=(f_1,\dots,f_s)$ and on the labeled node $a$, whose action on
networks is described in Fig.~\ref{fig:construction}.

\begin{figure}
\centering
\includegraphics[width=7cm]{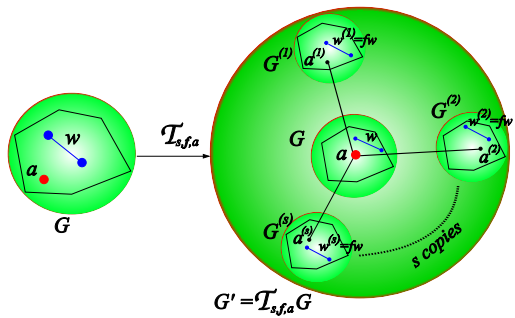}
\caption{{\it The map $\mathcal{T}_{s,\mathbf{f},a}$}. On the left a generic initial
  graph $G$ with its attaching node $a$ (red on-line) and a generic weighted
  edge $w\in G$ (blue on-line). On the right the new 
graph $G^{\prime}$ obtained as follows: Let $G^{(1)},\dots,G^{(s)}$ be $s$
copies of $G$, whose weighted edges (blue on-line) have
 been scaled respectively by a factor $f_1,\dots,f_s$, and let us denote by
 $a^{(i)}$, for $i=1,\dots,s$, the node in $G^{(i)}$ image of the labeled
 node $a\in G$, then link all those labeled nodes to $a\in G$ through edges of unitary weight. The connected
 network obtained in this way will be by definition the image of $G$ through the map: 
 $G^{\prime}=\mathcal{T}_{s,\mathbf{f},a}(G)$.}
\label{fig:construction}
\end{figure}

So starting with a given initial network $G_0$ we can construct a family of
weighted networks 
$(G_k)_{k\geq 0}$ iteratively applying the map $\mathcal{T}_{s,\mathbf{f},a}$:
$G_k=\mathcal{T}_{s,\mathbf{f},a}(G_{k-1})$.  

This construction improves the one recently proposed~\cite{carlettirighi} by
avoiding the introduction of an extra node, moreover it offers a unifying 
framework where several constructions presented in literature
can be included and generalized, e.g. the model presented in~\cite{JKK2002} 
with $m=3$ can
be mapped into to the WFN with $s=3$, $\mathbf{f}=(1,1,1)$, i.e. no weights, 
and $G_0=\bullet$. Finally this deterministic construction will be the basic 
brick to develop the stochastic network introduced in the following Section~\ref{sect:StochWFN}.

Given $G_0$ and the map $\mathcal{T}_{s,\mathbf{f},a}$ we 
are able to completely characterize the topology of each $G_k$ for $k\geq 1$ and also of the
limit network $G_{\infty}$, defined as the fixed point of the map,
$\mathcal{T}_{s,\mathbf{f},a}(G_{\infty})=G_{\infty}$.

\subsection{Results}
\label{sect:result}

 The aim of this section is to describe the topology of the graphs $G_k$ for
 all $k\geq 1$ and $G_{\infty}$, by analytically
studying their properties such as the average degree, the node
strength distribution, the average (weighted) shortest path and the (weighted)
clustering coefficient.  

At each iteration step the graph $G_k$ grows as the number of its nodes
increases according to 
\begin{equation}
  \label{eq:numnodes}
  N_k=(s+1)^kN_0\, ,
\end{equation}
being $N_0$ the
number of nodes in the initial graph, while the number of edges satisfies
\begin{equation}
  \label{eq:numedges}
  E_k=(s+1)^k(E_0+1)-1\, ,
\end{equation}
being $E_0$ the number of edges in $G_0$. Hence in the limit of
large $k$ the average degree is asymptotically given by 
\begin{equation}
  \label{eq:asymtavdeg}
  \frac{E_k}{N_k}\underset{k\rightarrow
    \infty}{\longrightarrow}\frac{E_0+1}{N_0}\, . 
\end{equation}

Let us denote the weighted degree of
node $i\in G_k$, also called {\it node strength}~\cite{BBPV2004}, by
$\omega^{(k)}_i=\sum_{j}w_{ij}^{(k)}$, being $w_{ij}^{(k)}$ the weight
of the edge $(ij)\in G_k$; then using the recursive construction,
we can explicitly compute the total node 
strength, $W_k=\sum_{i}\omega^{(k)}_{i}$, and easily show
that 
\begin{equation}
  \label{eq:wk}
  W_k=\left[\frac{2s}{F}\left((F+1)^k-1\right)+(F+1)^kW_0\right]\, ,
\end{equation}
being $F=\sum_{j=1}^s f_j$. Let us observe that using the hypothesis $f_j<1$,
it trivially follows that $F<s$, hence we
can conclude that the average node strength goes to zero as $k$
increases: ${W_k}/{N_k}\underset{k\rightarrow \infty}{\longrightarrow} 0$.

\subsection{Node strength distribution.}
\label{ssec:degdist}

 Let $g_k(x)$ denote the number of nodes in $G_k$ that have strength
 $\omega^{(k)}_i=x$ and let us assume $g_0$ to have values in some finite 
discrete subset of the positive reals, namely: 
\begin{equation*}
g_0(x)>0 \; \text{if and
  only if} \; x\in\{x_1,\dots,x_m\}\, ,
\end{equation*}
otherwise $g_0(x)=0$. Using the property of the map
$\mathcal{T}_{s,\mathbf{f},a}$ we get that after $k$ steps of the construction
the nodes strengths have been rescaled by a factor $f_1^{k_1}\dots f_s^{k_s}$,
where 
the non-negative integers $k_i$ do satisfy $k_1+\dots +k_s\leq k$. Because this
can be done in $k!/(k_1!\dots k_s!)$ possible different ways, we get the
following 
relation for the node strength distribution for the network $G_k$:
\begin{equation}
  \label{eq:nhnodestrength}
  g_k(f_1^{k_1}\dots f_s^{k_s}x)=\frac{k!}{k_1!\dots k_s!}g_0(x)\quad
  \text{with $k_1+\dots+k_s\leq k$}\, .
\end{equation}
After sufficiently many steps and assuming that the main contribution arises
from the choice $k_1\sim \dots \sim k_s \sim k/s$, we can use Stirling formula
to get the approximate distribution (see Fig.~\ref{fig:construction1bis})
\begin{equation}
  \label{eq:nhnodestrapprx}
  \log g_k(x) \sim \frac{s\log s}{\log (f_1\dots f_s)} \log x\, ,
\end{equation}
so the nodes strength distribution follows a power law. Let us observe that in
the case of homogeneous scaling, i.e. all $f_j$ equal to some $f\in (0,1)$,
one can prove~\cite{carlettirighi} that Eq.~\eqref{eq:nhnodestrapprx} reduces
to $\log g_k(x)\sim -d_{fract}\log x$ where $d_{fract}=-\log s/\log f$ is the
fractal dimension of the underlying IFS fractal.

\begin{figure}
\centering
\includegraphics[width=9cm]{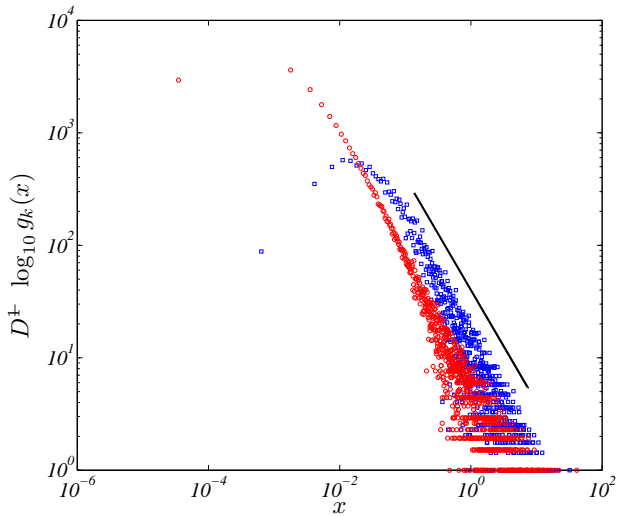}
\caption{{\it Node Strengths Distribution}. Plot of the renormalized node
  strengths 
  distribution $D^{-1}\log_{10}g_k(x)$, where $D=-s\log s/\log (f_1 \dots
  f_s)$. Symbols refer to : $\Box$ the finite approximation $G_{11}$ with
  $3145728$ 
  nodes of the WFN with $s = 3$, $\mathbf{f}
  =(1/\sqrt{2},1/\sqrt{3},1/\sqrt{5})$ and $G_0=\wfnone$;
  $\bigcirc$  the finite approximation $G_9$ with $3359232$ nodes of the WFN
   $s=5$, $\mathbf{f} =
  (1/\sqrt{5},1/\sqrt{11},1/\sqrt{3},1/\sqrt{7},1/\sqrt{13})$ and
  $G_0=\wfntwo$. The   
  reference line has slope $-1$, linear best 
  fits (data not shown) provides a slope $-1.037\pm 0.04$ and $R^2=0.798$ for
  $\Box$ and $-1.00\pm 0.03$ and $R^2=0.8382$ for $\bigcirc$.}  
\label{fig:construction1bis}
\end{figure}

\subsection{Average weighted shortest path.}
\label{ssec:meanpath}

By definition the average {\it weighted shortest path}~\cite{BLMCH2006} of the
graph $G_k$ is
\begin{equation}
\label{eq:wmpath}
 \lambda_k=\frac{\Lambda_k}{N_k(N_k-1)}\, ,
\end{equation}
 where
\begin{equation}
\label{eq:totwmpath}
\Lambda_k=\sum_{ij\in G_k} p_{ij}^{(k)}\, ,
\end{equation}
being $p_{ij}^{(k)}$ the weighted shortest path linking nodes $i$ and $j$ in 
$G_k$. Taking advantage of the
recursive construction and adapting the ideas used in~\cite{carlettirighi}, we get the following recursive relation for $\Lambda_k$
\begin{equation}
  \label{eq:Lkrec}
  \Lambda_k=(F+1)\Lambda_{k-1}+2s(F+1)N_{k-1}\Lambda_{k-1}^{(a_{k-1})}+2s^2N^2_{k-1}\,
  ,
\end{equation}
where we introduced $\Lambda_k^{(a_k)}=\sum_{i\in G_k}p_{ia_k}^{(k)}$,
i.e. the sum of all weighted shortest 
paths ending at the attaching node, $a_k\in G_k$. We can
prove that for large $k$ the asymptotic behavior of 
$\Lambda_k^{(a_k)}$ is given by
\begin{equation}
  \label{eq:Lkcasymt}
  \Lambda_k^{(a_k)}\underset{k\rightarrow \infty}{\sim}
  \frac{sN_0}{s-F}(s+1)^{k}\, ,
\end{equation}
and thus the recursive relation~\eqref{eq:Lkrec} can be explicitely solved
to provide the following asymptotic behavior 
in the limit of large $k$ (see Fig.~\ref{fig:renoweigmeanpath})
\begin{equation}
  \label{eq:ellkasym}
  \lambda_k\underset{k\rightarrow
    \infty}{\longrightarrow } 
  \frac{2s^2 (s+1)}{(s-F)[(1+s)^2-(1+F)]}\, . 
\end{equation}

 \begin{figure}
 \centering
 \includegraphics[width=8cm]{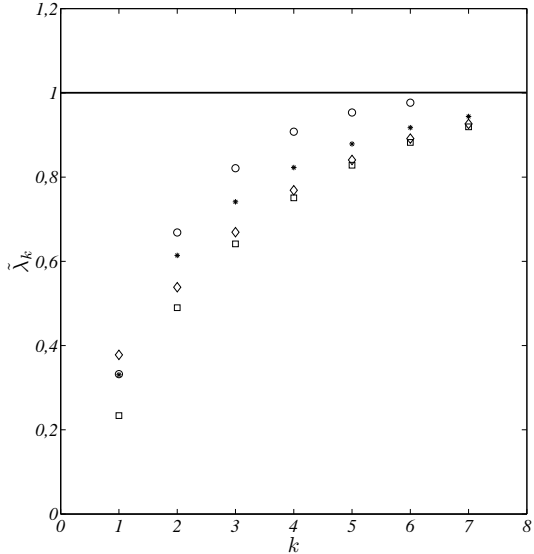}
\caption{{\it The average weighted shortest path}. Plot of the renormalized
  average 
  weighted shortest path $\tilde{\lambda}_k$ versus the iteration number $k$,
  where $\tilde{\lambda}_k=\lambda_k\frac{(s-F)[(1+s)^2-(1+F)]}{2s^2(s+1)}$
  and $F=f_1+\dots+f_s$. Symbols refer to : $\Box$ the WFN $s=3$,
  $\mathbf{f}=(1/\sqrt{2},1/\sqrt{3},1/\sqrt{5})$ and $G_0=\wfnone$;
  $\bigcirc$ the WFN $s=$5,
  $\mathbf{f}=(1/\sqrt{5},1/\sqrt{11},1/\sqrt{3},1/\sqrt{7},1/\sqrt{13})$ and
  $G_0=\wfntwo$; $\Diamond$ the WFN $s=2$,
  $\mathbf{f}=(1/\sqrt{3},1/\sqrt{5})$ and $G_0=\wfnthree$; $*$ the WFN
  $s=2$, $\mathbf{f}=(1/\sqrt{3},1/\sqrt{5})$ and $G_0=\wfnfour$.}
 \label{fig:renoweigmeanpath}
 \end{figure}

One can explicitly compute the {\it average shortest path}, $\ell_k$, formally
obtained by setting $f_1=\dots=f_s=1$ in the previous formulas~\eqref{eq:wmpath}
and~\eqref{eq:totwmpath}. Hence slightly 
modifying the results  
presented above we can prove that asymptotically we
have (see Fig.~\ref{fig:ellek})
\begin{equation}
  \label{eq:avmeanpath}
  \ell_k  \underset{k\rightarrow \infty}{\sim} \frac{2s}{(1+s)\log (s+1)}\log
  \frac{N_k}{N_0}\, ,  
\end{equation}
where growth law of $N_k$ given by~\eqref{eq:numnodes} has been used. Thus the network grows unbounded with
the logarithm of the network size, while the weighted shortest distances 
stay bounded.

\begin{figure}
\centering
\includegraphics[width=8cm]{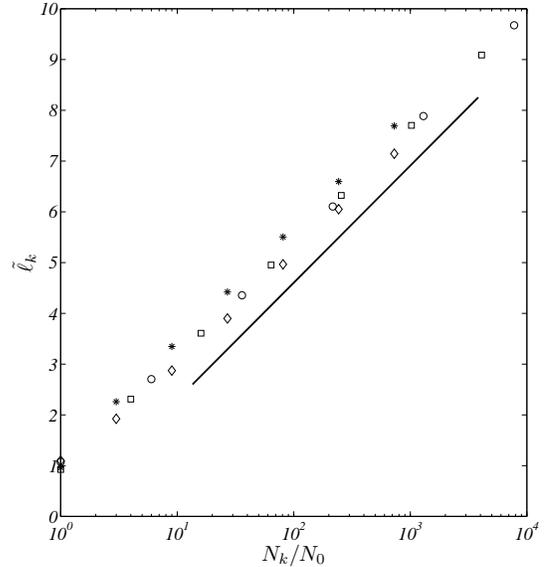}
\caption{{\it The average shortest path $\ell_k$ as a function of the network
    size} (semilog plot). Semilog plot of the renormalized average shortest
  path $\tilde{\ell}_k$ versus the network size $N_k$, where
  $\tilde{\ell}_k=\ell_k \frac{(s+1)\log (s+1)}{2s}$. 
Symbols are the same of Fig.~\ref{fig:renoweigmeanpath}. The reference line
has slope $1$.  
 Linear best fits (data not shown) provides a slope $0.970\pm 0.017$ and
  $R^2=0.9999$ for $\Box$, $0.9654 \pm 0.05$ and $R^2=0.9997$ for
  $\bigcirc$, $0.97\pm 0.02$ and $R^2=0.9998$ for $\Diamond$ and $1.01\pm
  0.03$ and $R^2=0.9993$ for $*$.}
\label{fig:ellek}
\end{figure}

\subsection{Clustering coefficient.}
\label{ssec:cluscoeff}
 The clustering coefficient~\cite{WattsStrogatz1998,BLMCH2006} of the graph
 $G_k$ is defined as the average over the whole set of
 nodes of the local clustering coefficient $c_i^{(k)}$, namely $<c_k>=C_k/N_k$,
 where $C_k=\sum_{i\in G_k}c^{(k)}_i$. 

Because of the 
construction algorithm new triangles are created in the network \lq\lq
boundary\rq\rq while their number doesn't change in the inner core, hence the local clustering coefficient, at
each step increases just by a factor $s+1$; thus after $k$--interactions we will have
 $C_k=(1+s)^{k}C_0$, being $C_0=\sum_{i\in G_0}c^{(0)}_i$ the sum of
 local clustering coefficients in the initial graph. We can thus conclude that
 the clustering 
 coefficient of the graph is asymptotically 
 given by: 
 \begin{equation}
   \label{eq:asymptclustcoeff}
   <c_k>\underset{k\rightarrow \infty}{\longrightarrow }
   \frac{C_0}{N_0}\, . 
 \end{equation}

 On the other hand, one can introduce the links values to weigh the clustering
 coefficient~\cite{SKOKK2007}, generalizing the previous relation, we can
 easily prove that {\it weighted clustering coefficient} of the graph is
 asymptotically given by: 
 \begin{equation}
   \label{eq:asymptclustcoeff}
   <\gamma_k>=\left(\frac{1+F}{1+s}\right)^k\frac{C_0}{N_0} \underset{k\rightarrow \infty}{\sim}
   \frac{1}{N_k^{1-d}}\, ,
 \end{equation}
where $d=\frac{\log (1+F)}{\log (1+s)}$, that results smaller than one because
of the assumption $f_j<1$.

\section{Stochastic Weighted Fractal Networks}
\label{sect:StochWFN}

The aim of this section is to present a class of complex weighted networks that grow according to a {\it stochastic} process and exhibit self-similar or self-affine structures, hereby named {\it Stochastic Weighted Fractal Networks}, for short {\it SWFN}, whose construction is directly inspired by the stochastic growth phenomena present in nature. The idea is
thus to mimic the growth of fractal structures in nature where \lq\lq
possible errors\rq\rq could modify regular patterns.

So let us hypothesize that the growth process is the result of a
stochastic process that selects the actual realization, i.e. the number of copies, between a number of
different possibilities. Thus at each iteration the number of 
copies, $s$, is a stochastic variable distributed according to some probability distribution function $p(s)$. Once the numerical value for $s$ has been set, $s$ real numbers $f_1,\dots,f_{s}$ are drawn according to some probability
distribution function $q(f)$ with values in $(0,1)$. Finally a new network is constructed
by applying $\mathcal{T}_{s,(f_1,\dots ,f_{s}),a}$ to the actual network:
\begin{equation}
  \label{eq:stoc1b}
G \underset{p(s)}{\longmapsto} G^{(s)}=\mathcal{T}_{s,(f_1,\dots
  ,f_{s}),a}G \, . 
\end{equation}

\medskip
{\bf Remark.} \label{rem:fassump}{\it In the following we will assume the simplifying working
hypothesis that $f_1=\dots =f_{s}=\alpha/s$, i.e. $q(f)=\delta(f-\alpha/s)$, for some given and fixed $\alpha
\in (0,1)$, but of course the model applies to more 
general cases.}

\medskip
One can repeat the construction $k$ times and thus obtain with
probability $p(s_k)\dots p(s_1)$, starting
from a network $G_0$, a new network, denoted by $G^{(s_k, \dots , s_1)}$:
\begin{equation}
  \label{eq:kstepstoc1b}
G^{(s_k, \dots , s_1)}=
\mathcal{T}_{s_k,(f^{(k)}_1,\dots
  ,f^{(k)}_{s_k}),a}\circ \dots \circ \mathcal{T}_{s_1,(f^{(1)}_1,\dots
  ,f^{(1)}_{s_1}),a} G_0\, .
\end{equation}
The network growth results thus a stochastic process, hence we will
describe the main topological network measures in terms of {\it expectations}
obtained 
repeating several times the construction. Of course we
could also consider and compute higher order momenta, but the computations
become 
rapidly cumbersome, and thus we will non present these results except for
some simple cases, such as the number of nodes.

\subsection{Results: SWFN}
\label{ssec:resswfn-s}

At each step the number of nodes increases with respect the present ones, and
the exact amount depends on the number of branches drawn. Starting 
from a network containing $N_0$ nodes we get a new network with
$N^{(s_1)}=(1+s_1)N_0$ nodes with probability $p(s_1)$. Iterating the
construction, after $k$ steps we can obtain with probability
$p(s_k)\dots p(s_1)$ a network with $N^{(s_k, \dots , s_1)}=(1+s_k)\dots
(1+s_1)N_0$ nodes. Hence the expected value for the number of nodes in a
network build after $k$ iterations, is given by:
\begin{eqnarray}
  \label{eq:expnodes}
 &\phantom{x}& <N_k>=\sum_{s_k,\dots,s_1}p(s_k)\dots p(s_1)N^{(s_k, \dots , s_1)}\\ 
 &=&\sum_{s_k}p(s_k) (1+s_k)\sum_{s_{k-1},\dots,s_1} p(s_{k-1})\dots p(s_1)
   N^{(s_{k-1}, \dots , s_1)}\notag \\
&=& (1+<s>)<N_{k-1}>\notag \, ,
\end{eqnarray}

where we denoted by $<s>=\sum_{s_k}p(s_k)s_k$ the average number of branches. We can thus conclude that the
expected number of nodes increases exponentially
\begin{equation}
  \label{eq:nodes}
  <N_k>=(1+<s>)^kN_0\, .
\end{equation}
Using similar ideas one can prove that the variance of the number of nodes
increases according to:
\begin{equation}
  \label{eq:variancenodes}
  \sigma_{N_k}^2=N_0^2\left[ (1+<s>)^2+\sigma^2_s\right]^k-(1+<s>)^{2k}N_0\, ,
\end{equation}
where $\sigma_s^2$ is the variance of the distribution of number of branches.

On the other hand the number of edges can increase, with probability $p(s_k)$, in one
iteration by $E^{(s_{k}, \dots , s_1)}=(1+s_k)E^{(s_{k-1}, \dots , s_1)}+s_k$
and thus  
the expected number of edges do satisfy
\begin{equation}
  \label{eq:edges}
  <E_k>=(1+<s>)^k(E_0+1)-1\, .
\end{equation}
These findings are exact in the case of infinitely many replicas, nevertheless
numerical simulations presented in Fig.~\ref{fig:expNEavdegavstr} and in
Fig.~\ref{fig:expWavdegavstr} show the
good agreement also for finitely many repetitions.

\smallskip
{\bf Remark.} {\it The numerical simulations presented in the following will
  be obtained assuming for the branch number a Poisson distribution
  translated by one, more precisely to avoid a non zero probability of drawing zero branches, we
  drawn with probability $p(k)=\lambda^k e^{-\lambda}/k!$ a non negative
  integer $k$, and then we set the number of branches to $s=k+1$, in this way
  we will get $<s>=\lambda+1$, $\sigma^2=\lambda$ and $s\geq 1$.

Of course our findings are more general and do not rely on the particular
choice for $p(s)$.}

\begin{figure*}
\centering
\includegraphics[width=8cm]{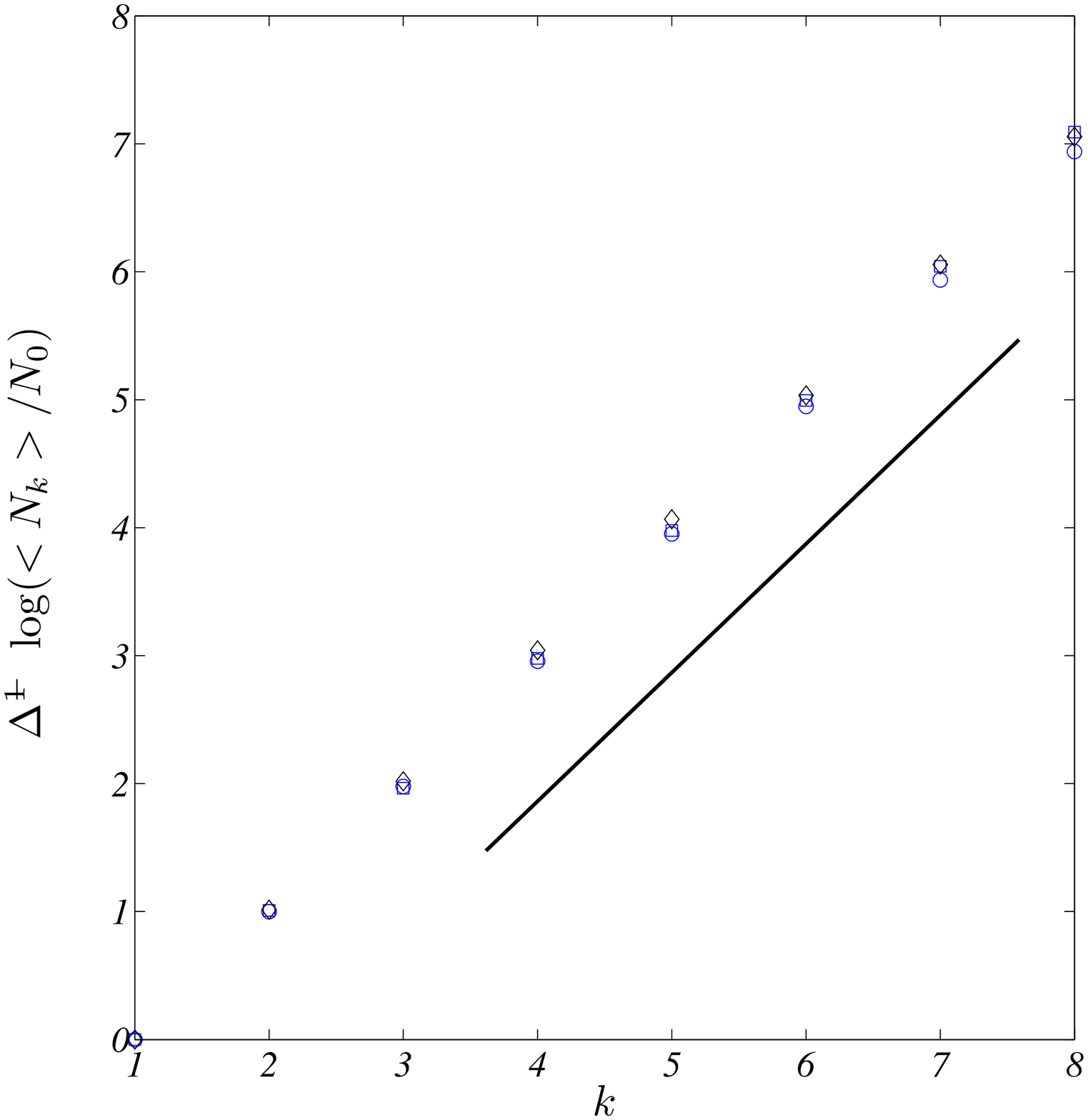}\quad
\includegraphics[width=8cm]{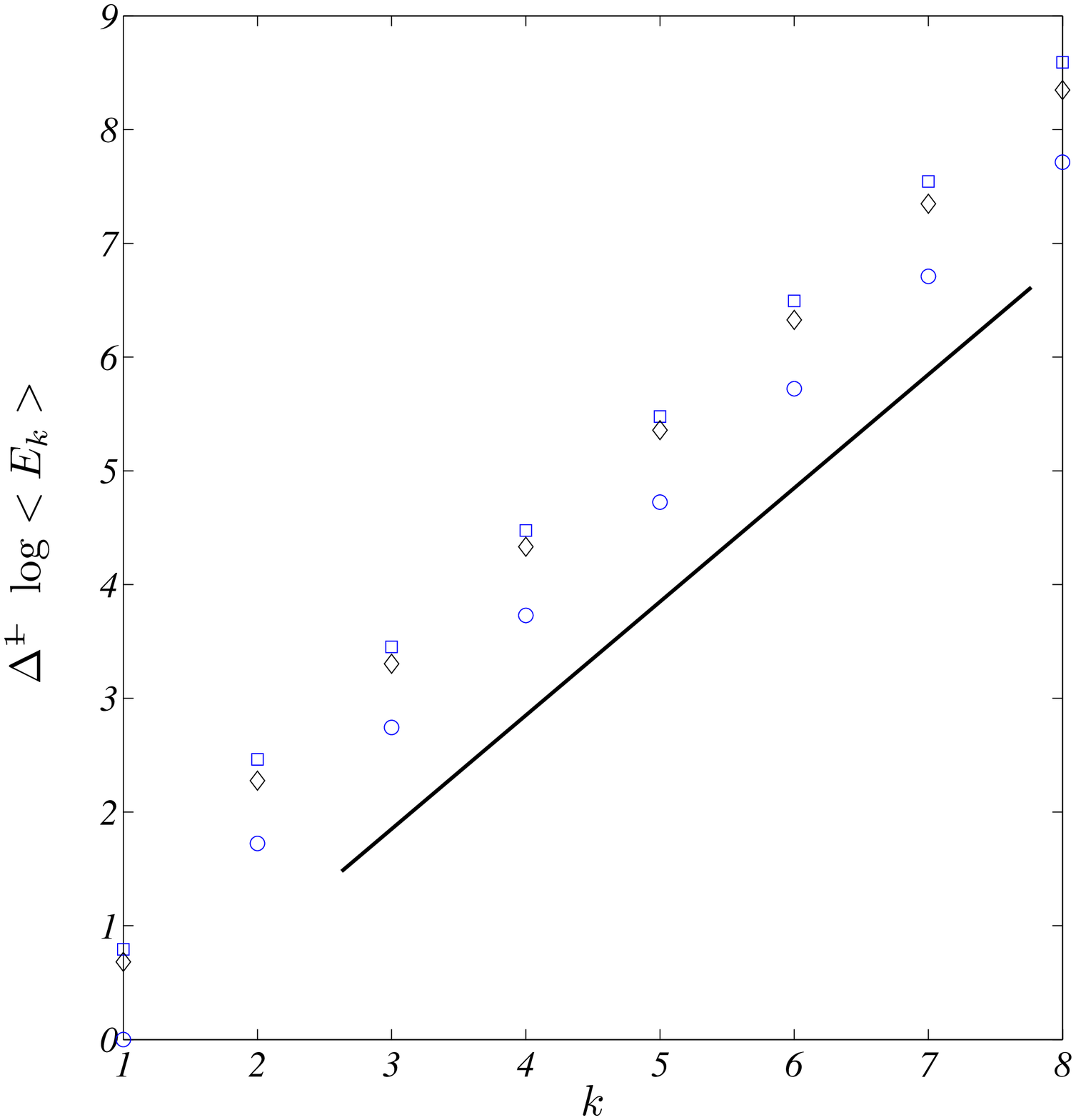}
\caption{{\it Expected values for number of Nodes and number of Edges}.
  Renormalized quantities : $\Delta^{-1}\log (<N_k>/N_0)$ and
  $\Delta^{-1}\log <E_k>$ where 
  $\Delta=\log(1+<s>)$. Symbols refer to : $\bigcirc$ the SWFN with parameters
  $\lambda=4$, $\alpha=0.5$ and $G_0=\wfntwo$; $\Box$ the SWFN with parameters
  $\lambda=2$, $\alpha=0.5$ and $G_0=\wfnone$; $\Diamond$ the SWFN with
  parameters $\lambda=3$, $\alpha=0.8$ and $G_0=\wfnone$.
Expectations are obtained over $100$ replicas. Left panel, the reference
line has slope $1$, linear best fits (data not shown) give $0.9998\pm
0.03$ $R^2=0.9991$ for $\bigcirc$ and $1.017\pm 0.008$ $R^2=0.9999$ for
$\Box$, $1.008\pm 0.005$ $R^2=1.000$ for $\Diamond$. Right panel, the reference
line has slope $1$, linear best fits (data not shown) give $0.9569\pm 0.06$
$R^2= 0.9988$ for $\bigcirc$, $1.019\pm 0.03$ $R^2=0.9997$ for $\Box$ and
$1.06\pm 0.07$ $R^2=0.9955$ for $\Diamond$.} 
\label{fig:expNEavdegavstr}
\end{figure*}

\smallskip
In a similar way we can compute the expected average degree after $k$ steps,
$<(E/N)_k>$, and 
the expected average node strength after $k$ steps, $<(W/N)_k>$, where $W$ is
the total node strength for the given network realization, to get 
(see Fig.~\ref{fig:expWavdegavstr}): 
\begin{equation}
\label{eq:expavdeg}
\left< \left(\frac{E}{N}\right)_k\right>\underset{k \rightarrow
  \infty}{\rightarrow} \frac{E_0+1}{N_0}
\quad \text{and} \quad \left< \left(\frac{W}{N}\right)_k\right>\underset{k
  \rightarrow 
  \infty}{\rightarrow} 0\, .
\end{equation}

\begin{figure*}
\centering
\includegraphics[width=8cm]{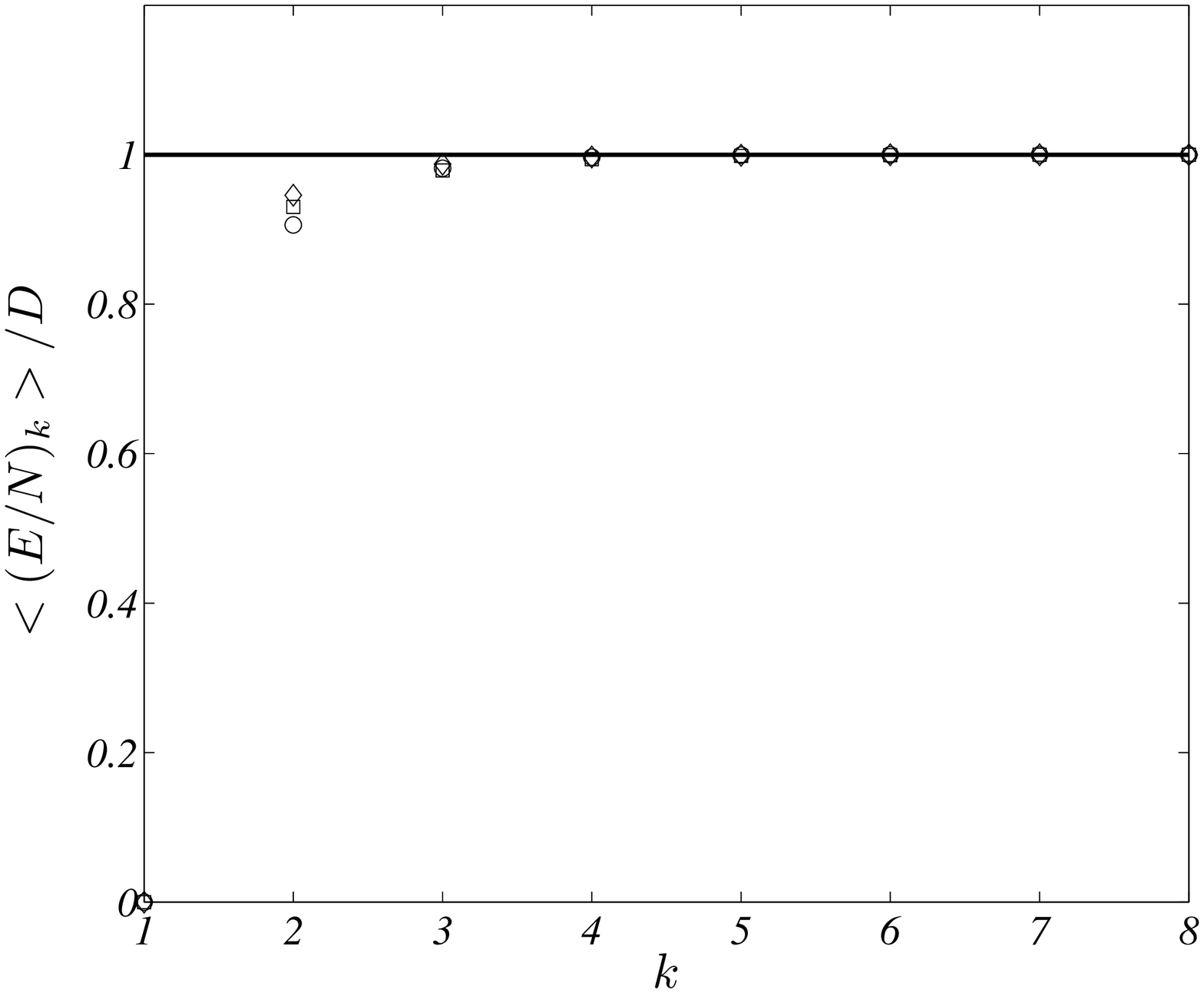}\quad
\includegraphics[width=8cm]{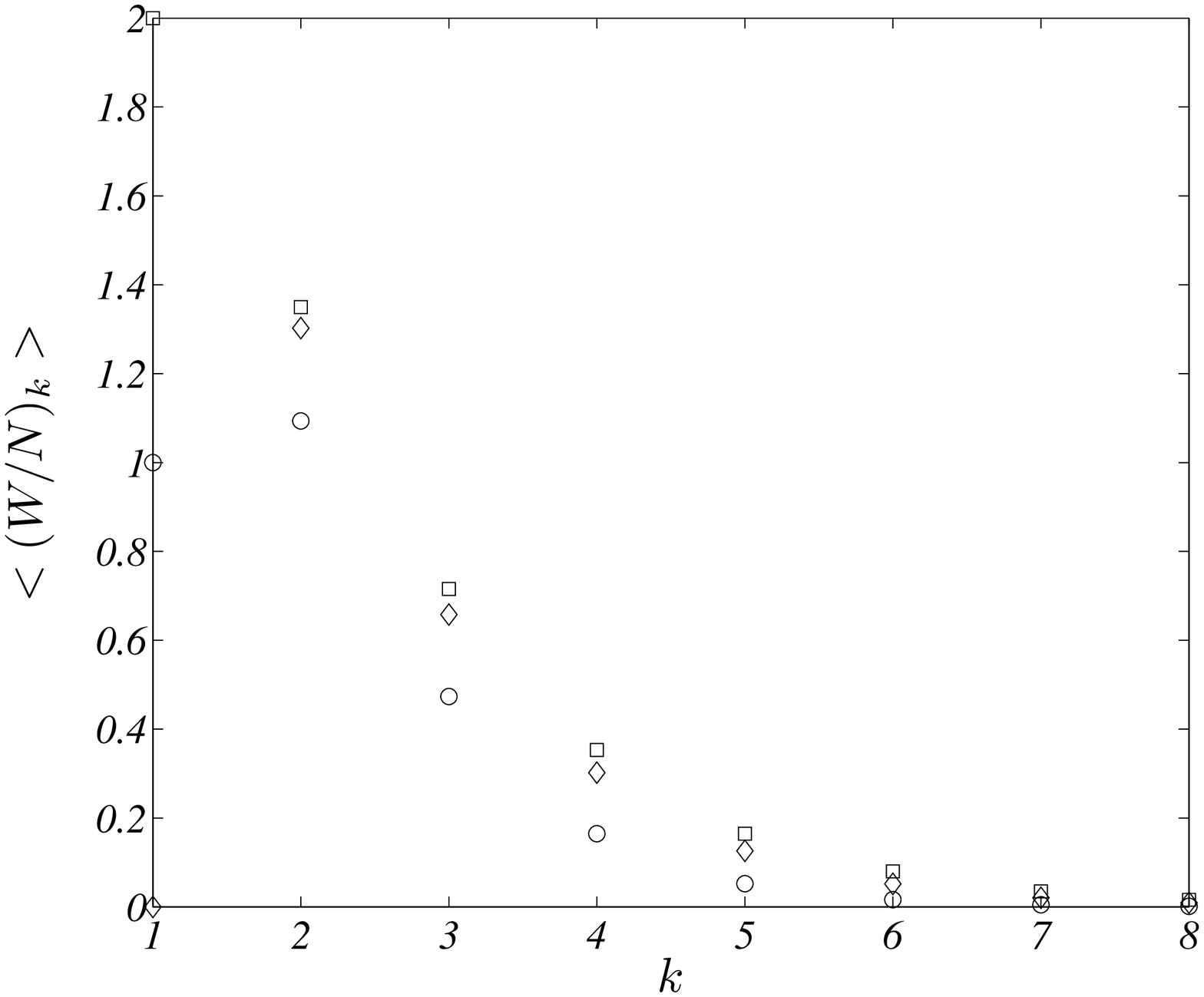}
\caption{{\it Expected values for the average
  degree and the average node strength}. Renormalized quantities :
  $<(E/N)_k>/D$ where $D=(E_0+1)/N_0$. Symbols are the same of
  Fig.~\ref{fig:expNEavdegavstr}. Expectations made over $100$ replicas.}
\label{fig:expWavdegavstr}
\end{figure*}

As we did in the previous section, we are able to analytically study other
relevant quantities such as the {\it expected} value for the {\it weighted shortest 
path} $<\lambda_k>$, defined for each network realization
by~\eqref{eq:wmpath}. More precisely, starting from a network $G_0$ and
applying iteratively the above construction 
we end up after $k$ iterations with
probability $p(s_k)\dots p(s_1)$ to a network $G^{(s_k,\dots ,s_1)}$, we can
thus define the weighted shortest path for the given network realization
by $\lambda^{(s_k,\dots  
  ,s_1)}= \frac{\Lambda^{(s_k,\dots ,s_1)}}{\left(N^{(s_k,\dots
      ,s_1)}\right)^2}$. Then using the recursive construction we get: 
\begin{widetext}
  \begin{eqnarray}
  \label{eq:lambdaGk}
  \lambda^{(s_k,\dots ,s_1)}&=& \frac{(F_k+1) \Lambda^{s_{k-1},\dots
      ,s_1}+2s_k^2 \left(N^{(s_{k-1},\dots 
      ,s_1)}\right)^2}{(1+s_k)^2\left(N^{(s_{k-1},\dots
      ,s_1)}\right)^2}+\frac{2s_k(F_k+1) \Lambda_{a_k}^{s_{k-1},\dots
    ,s_1}N^{(s_{k-1},\dots ,s_1)}}{(1+s_k)^2
  \left(N^{(s_{k-1},\dots ,s_1)}\right)^2}\notag \notag \\
&=& \frac{F_k+1}{(1+s_k)^2}\lambda^{(s_{k-1},\dots ,s_1)}
+2\left(\frac{s_k}{s_k+1}\right)^2+\frac{2s_k(F_k+1)}{(1+s_k)^2}\hat{\lambda}^{(s_k,\dots
  ,s_1)}\, ,
\end{eqnarray}
\end{widetext}
where we used the growth rate of the number of nodes and~\eqref{eq:Lkrec} and
we introduced $\hat{\lambda}^{(s_k,\dots ,s_1)}~=~\Lambda_{a_k}^{(s_{k},\dots
  ,s_1)}/N^{(s_{k},\dots ,s_1)}$. 

One can finally prove that the expected value for the
average weighted shortest path satisfies the recurrence equation:
\begin{widetext}
  \begin{equation}
  \label{eq:recexpwsp}
  \left< \lambda_k\right>=\left< \lambda_{k-1}\right>\left<\frac{1+F}{(1+s)^2}\right>
  +2\left< \left(\frac{s}{s+1}\right)^2\right> +2\left<
    \frac{s(1+F)}{(1+s)^2}\right>\left< \hat{\lambda}_k\right> \, ,
\end{equation}

\end{widetext}
where we defined
\begin{eqnarray}
  \label{eq:recexpwsphat}
  \left<\frac{1+F}{(1+s)^2}\right>&=&\sum_k p(k) \frac{1+F_k}{(1+s_k)^2} \, ,\notag \\
 \left< \left(\frac{s}{s+1}\right)^2\right>&=&\sum_k
  p(k)\left(\frac{s_k}{s_k+1}\right)^2 \, \text{ and }\notag \\
\left<
    \frac{s(1+F)}{(1+s)^2}\right>&=&\sum_k p(k)\frac{s_k(1+F_k)}{(1+s_k)^2}\, .
\end{eqnarray}


Under the simplifying assumption $f_1=\dots=f_{s_k}=\alpha/s_k$ we get
$F_k=\alpha$ and thus we 
can simplify the previous equations and obtain (see Fig.~\ref{fig:avwpath}):

\begin{widetext}
  \begin{equation}
  \label{eq:expwspanal2}
  <\lambda_k> \underset{k \rightarrow \infty}{\rightarrow}
  \left<\left(\frac{s}{s+1}\right)^2\right>\frac{2}{1-(\alpha+1)\left<1/(1+s)^2\right>}+\left<\frac{s}{(1+s)^2}\right>\left<\frac{s}{1+s}\right>\frac{2(\alpha+1)}{1-(\alpha+1)\left<1/(1+s)^2\right>}
\frac{1}{1-(\alpha+1)\left<1/(1+s)\right>}\,  .
\end{equation}
\end{widetext}

\begin{figure}
\centering
\includegraphics[width=8cm]{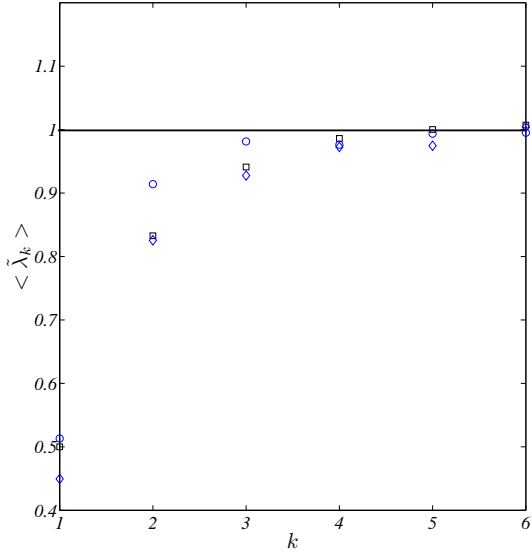}
\caption{{\it Expected values for the average weighted shortest 
    path}. Renormalized 
  quantities: $<\tilde{\lambda}_k>= L^{-1}<\lambda_k>$, where $L$ is the
right hand side of Eq.~\eqref{eq:expwspanal2}.
Symbols are the same of Fig.~\ref{fig:expNEavdegavstr}. Expectations made over
$20$ replicas.} 
\label{fig:avwpath}
\end{figure}

One can consider the {\it expected shortest path} by formally set all the scaling
factors equal to $1$ and similar technics allow to conclude that (see
Fig.~\ref{fig:avpath}) 
\begin{equation}
  \label{eq:expshrtpath}
  <\ell_k> \underset{k \rightarrow \infty}{\sim}\left<
    \left(\frac{s}{s+1}\right)^2\right>\frac{2}{1-\left<1/(s+1)\right>}\frac{1}{\log
    (1+<s>)}\log\frac{<N_k>}{N_0}\, .  
\end{equation}


{\bf Remark.} {\it 
Let us observe that in the case where only one value of $s$ is possible,
i.e. the probability distribution of the number of branches reduces to a
$\delta$--distribution, $p(s)=\delta_{s,s'}$, 
then the above result coincide with the ones presented for the WFN in
Section~\ref{sect:nHWFN}.} 

\begin{figure}
\centering
\includegraphics[width=8cm]{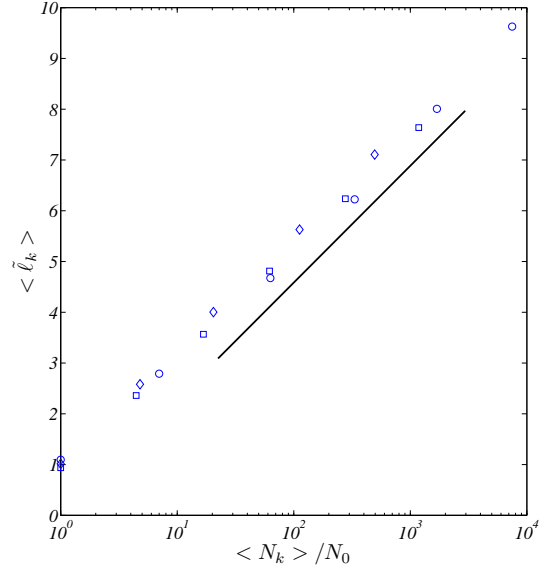}
\caption{{\it Expected values for the average shortest path as a function of the
    network size} (semilog plot). Semilog plot of the renormalized expected
  average shortest path $<\tilde{\ell}_k>=<\ell_k>/M$ versus the network size
  $N_k$, where 
  $M$ is the right hand side of Eq.~\eqref{eq:expshrtpath}.
Symbols are the same of Fig.~\ref{fig:expNEavdegavstr}. Expectations made over $20$ replicas. The reference line
has slope $1$, linear best fits (data not shown) provides $0.95\pm 0.01$ with
$R^2=0.9999$ for $\Box$, $0.95\pm 0.07$ $R^2=0.9968$ for $\bigcirc$ and
$0.98\pm 0.01$ $R^2=0.9999$ for $\Diamond$.}
\label{fig:avpath}
\end{figure}

\section{Conclusions}
\label{sect:conclusion}

In this paper we proposed a unifying general framework for complex weighted networks sharing several properties with fractal sets, the {\it Stochastic Weighted Fractal Networks}. This theory, that generalizes to networks the construction of physical fractals, allows us to build complex networks with a prescribed topology, whose main quantities can be analytically predicted in terms of expectations and have been
shown to depend on the fractal dimension of some underlying fractal; for instance the networks are scale-free, the exponent being the related to the fractal dimension of the underlying IFS. Moreover the SWFN share with fractals, the self-similar or self-affine structure.

These networks exhibit the small-world property. In fact the average shortest path increases logarithmically with the system size; hence it is as small as the average shortest path of a random network with the same number of nodes and same average degree. On the other hand the clustering coefficient is asymptotically constant, thus larger than the clustering coefficient of a random network that shrinks to zero as the system size increases.

As already observed~\cite{carlettirighi} the self-similarity property of the SWFN make them suitable to model real problems involving some kind of diffusion over the network coupled with local looses of flow, here modeled via the parameters
$f<1$. Moreover the stochastic growth process allows us to introduce  more realism in the construction.

\end{document}